\documentclass[nofootinbib,aps,prl,twocolumn,superscriptaddress,groupedaddress]{revtex4}  
\usepackage{graphicx}  
\usepackage{dcolumn}   
\usepackage{bm}        
\usepackage{amssymb}   

\def\beq{\begin{eqnarray}}
\def\eeq{\end{eqnarray}}

\def\({\left(}
\def\){\right)}

\def\mpl{M_{\rm pl}}

\def\f{\varphi}

\newcommand{\be}{\begin{equation}}
\newcommand{\ee}{\end{equation}}
\newcommand{\la}{\langle}
\newcommand{\ra}{\rangle}
\def\ea{\end{eqnarray}}
\def\ba{\begin{eqnarray}}

\def\beq{\begin{eqnarray}}
\def\eeq{\end{eqnarray}}

\def\({\left(}
\def\){\right)}

\def\mpl{M_{\rm Pl}}

\def\la{\langle}
\def\ra{\rangle}

\def\lsim{\mathrel{\rlap{\lower3pt\hbox{\hskip0pt$\sim$}}
     \raise1pt\hbox{$<$}}}         
\def\gsim{\mathrel{\rlap{\lower4pt\hbox{\hskip1pt$\sim$}}
     \raise1pt\hbox{$>$}}}         

\def\lsim{\mathrel{\rlap{\lower3pt\hbox{\hskip0pt$\sim$}}
     \raise1pt\hbox{$<$}}}         
\def\gsim{\mathrel{\rlap{\lower4pt\hbox{\hskip1pt$\sim$}}
     \raise1pt\hbox{$>$}}}         

\usepackage{amsmath}
\usepackage{amsfonts}
\usepackage{verbatim}
\usepackage{graphicx}
\usepackage{subfigure}
\usepackage{amssymb}
\usepackage[T1]{fontenc}
\usepackage{slashed}
\usepackage{color}

\begin{document}

\title{A Relativistic Gas of Inflatons as an Initial State for Inflation}
\author{Lasha Berezhiani}
\address{Max-Planck-Institut f\"ur Physik, F\"ohringer Ring 6, 80805 M\"unchen, Germany}
\author{Mark Trodden}
\address{Center for Particle Cosmology, Department of Physics and Astronomy, University of Pennsylvania,\\ 209 South 33rd St, Philadelphia, PA 19104}
\date{\today}

\begin{abstract}

The possibility of the resilience of the beginning of inflation under unfavorable conditions is examined by considering the initial state of the inflaton field to be in the form of a relativistic gas with some of its properties in close proximity to the black body spectrum. It is demonstrated that the initial potential energy budget in such an environment is suppressed beyond the minimal value required for inflation. This is the extension of our earlier work, where we have shown that the rare regions which happen to host favorable initial conditions for the beginning of inflation could come to dominate the late-time Universe only if they started out from above the so-called self-reproduction threshold. 
\end{abstract}
\maketitle

\section*{Introduction}

Cosmic Inflation provides a mechanism for producing the entire observable Universe from a single causal patch, many orders of magnitude smaller than the current Hubble radius \cite{Guth:1980zm,Linde:1981mu,Albrecht:1982wi}. By now it is straightforward to construct models in which an initial ultra-cold homogenous (potential energy dominated) and featureless Hubble patch evolves into a macroscopic universe which is in spectacular agreement with observations. An interesting question around the inflationary paradigm concerns the origin of this initial patch. Much effort has been invested in understanding how much (if any) fine-tuning of the initial conditions is required for inflation to begin~\cite{Hartle:1983ai,Linde:1985ub,Brandenberger:1990wu,Linde:1993xx,Farhi:1986ty,Farhi:1989yr,Vachaspati:1998dy,Kaloper:2002cs,Albrecht:2004ke,Gibbons:2006pa,Hartle:2008ng,Schiffrin:2012zf,Ijjas:2013vea,Guth:2013sya,Linde:2014nna,Ijjas:2014nta,Carroll:2014uoa,Mukhanov:2014uwa,East:2015ggf,Kleban:2016sqm,Berezhiani:2015ola,Aurrekoetxea:2019fhr,Clough:2017efm,Clough:2016ymm}. A tantalizing thought is that inflation is highly resilient and that it will begin under arbitrary initial conditions. Barring an additional special mechanism for producing preferable initial state, there are two possible scenarios for starting inflation in a universe with generic initial conditions:

\begin{itemize}

\item[{\bf i)}] Even if regions with conditions favouring the beginning of inflation are initially rare, they are rewarded with an exponential volume factor due to accelerated expansion. Thus, they still may come to dominate the physical volume of the universe.

\item[{\bf ii)}] In spite of a given region being highly inhomogeneous, it may evolve into one capable of seeding inflation if, in the expanding universe, all forms of energy redshift away quickly, except for the slowly-rolling potential energy.

\end{itemize}

There are a number of different techniques for investigating these questions, including, most popularly, simulating given inflationary models with varying initial conditions in order to understand the evolution towards inflation in the spirit of option (ii). Recently a significant progress has been made in this direction \cite{East:2015ggf,Clough:2016ymm,Clough:2017efm,Aurrekoetxea:2019fhr}. There one usually considers highly inhomogeneous initial inflaton profile of the form
\beq
\f=\f_0+\delta\f \,.
\eeq
The constant $\f_0$ is chosen to be capable of giving sufficient inflation if one were to remove inhomogeneities. The latter component $\delta\f $ is considered to average to zero over the simulation box; which is chosen to be of the size of the curvature radius sourced by $\f_0$ only. The simulations are performed for quite a broad (with $|\delta\f |\gg\phi_0$), yet limited, class of classical inhomogeneities, demonstrating that for such initial conditions inflation persists. However, if $\f_0$ is significantly smaller than the above-mentioned critical value, the same simulations show that the outcome is not as favourable.

In previous work~\cite{Berezhiani:2015ola}, we have taken a complementary analytic approach, in which we describe the initial state in terms of constituent quanta. In order to quantify the probability distribution for the pre-inflationary universe, one then has to make assumptions about the production mechanism. We considered the possibility that the inflaton quanta were produced through high energy processes, and used this technique to explore the option (i) for the onset of inflation. In this way we demonstrated that regions smooth enough to seed inflation are quite rare, so that even the exponential volume factor is insufficient to overcome it, unless one enters the regime of slow-roll eternal inflation. This underlined the importance of gaining an improved understanding of this stage of evolution. This approach was motivated by the corpuscular description of inflation developed in \cite{Dvali:2013eja}, where classical field configurations are represented as a coherent state of inflatons and gravitons. Semi-classically, the quantum dynamics of a spectator scalar field on a fixed de Sitter background is considered as a fair proxy for eternal inflation, in which case the question of the beginning of inflation is rendered far less pressing. However, it has also been argued \cite{Dvali:2013eja} that the fundamental treatment of classical backgrounds such as quasi-de-Sitter space may well be in terms of the coherent state built around the well-defined (Minkowski) vacuum. One of the justifications for this picture is the S-matrix formulation of quantum gravity \cite{Dvali:2020etd}. Using this method one concludes that there is a significant loss of coherence for regions that start out deep within the slow-roll eternal inflationary regime.

The goal of this paper is to extend the arguments of~\cite{Berezhiani:2015ola}, in the context of option (ii). Our work is intended to complement the results obtained through numerical simulations, by assessing whether a sufficiently homogeneous component $\f_0$, which is the input for those simulations, is expected to be present under the assumption that the initial state is in the form of the relativistic inflaton gas. It is clear from the very beginning that our assumptions deliberately suppress the initial potential energy budget, i.e. $\f_0$. The point we would like to understand is by how much.

To reiterate the concept, we are probing the resilience of inflation by pinpointing the types of initial conditions for which its onset is inefficient. Moreover, we do so by considering the initial state to have properties that one encounters in actual physical systems. For instance, as is well known, a gas of relativistic particles (with vanishing chemical potential) in a black body cavity at temperature $T$ obeys Planck's distribution per unit volume
\beq
dn_k\propto\frac{k^2dk}{e^{k/T}-1}\,,
\eeq
with $\int dn_k\propto T^3$.

Our proposal is to consider initial states of the inflaton gas having properties resembling this well known system, so that the expected value of the initial potential energy budget, and correspondingly $\f_0$,  can be estimated as the energy density of nonrelativisitic particles. This paper is dedicated to analyzing the consequences of such assumptions.

\section*{The Basic Idea} 

For simplicity, let us focus on a simplest model, in which the inflaton is a non-self-interacting massive scalar field, minimally coupled to gravity\footnote{Although it is disfavoured by observations, this model represents a convenient playground for exploring theoretical questions.}
\beq
\mathcal{L}_\varphi=\sqrt{-g}\left( -\frac{1}{2}g^{\mu\nu}\partial_\mu \varphi \partial_\nu \varphi- \frac{1}{2}m^2\varphi^2 \right)\, ,
\eeq
with the mass $m$ significantly smaller than the energy and curvature scales of interest.

The approach proposed in \cite{Berezhiani:2015ola}, was to treat the initial state of $\varphi$ as we would treat that of any other particle - namely to consider its initial configuration as being comprised of $\varphi$-quanta defined around the true vacuum of the theory, which we assume to be Minkowski space, setting the Cosmological Constant to zero for simplicity. Obviously, the collection of such quanta will not always behave as a quasi-ideal gas of on-shell particles, especially in the setting of the early universe. In fact, depending on the density and momentum distribution they could be significantly off-shell in some situations (see e.g. \cite{Berezhiani:2016grw}).  We will estimate how many on-shell inflaton particles with certain kinematic properties should be produced in a given region in order for them to go off-shell and collectively behave as a desirable classical configuration.

In order to make any useful statement, it is necessary to have some idea about the  mechanism through which the field is produced. We choose to do this by considering the production of $\varphi$ degrees of freedom through perturbative high-energy scattering and decay processes of some relativistic ultra-violet (UV) degrees of freedom. This implies that quanta would be predominantly produced in high momentum states, and we account for this by assuming that in a region with energy density $\rho$ it is unlikely for a given inflaton quantum to be much softer than the characteristic inter-particle separation. We quantify this suppressed probability as
\beq
p(k<k_0)=\left( \frac{k_0}{E} \right)^\alpha\,,
\label{prob1}
\eeq
with $\alpha\geq 1$, for $k_0\ll E\equiv\rho^{1/4}$. Moreover, we assume that the corresponding probability distribution contains a single energy scale $E$ and obeys $\la k \ra\simeq E$. This behavior is satisfied by the black body spectrum outlined at the end of the previous section, but is also true of a more general family of distributions.

The key point here is that we would like to consider an initial state in which inflatons behave as radiation on average and thus are not predisposed to aid the beginning of inflation. We then study the resilience of inflation in this environment; one not designed ahead of time to ensure accelerated stage of expansion. Notice that \eqref{prob1} leads to the suppression of the homogeneous component of the field configuration, and correspondingly of the potential energy budget in the initial state. Taking into account that a priori we have no information about the pre-inflationary universe, it is vital to look into such initial states to have a full understanding of the question of robustness of the paradigm. Such challenges are important for understanding just how strong of an attractor we are dealing with, and to inform us about what we may need to further understand about the dynamics of the universe before the observable stage\footnote{This is related to the idea \cite{Aurrekoetxea:2019fhr} that inflation can be a test of fundamental theories not only through its constraints on dynamical models, but also via constraints on the space of initial conditions.}.

For example, it was demonstrated in \cite{Berezhiani:2015ola} that within the corpuscular approach to the initial state (even requiring that the homogeneous component is suppressed) the problem could be ameliorated if slow-roll could be extended beyond the self-reproduction threshold. In that case it was not even necessary to appeal to the potentially infinite inflated volume predicted by the semi-classical analysis of eternal inflation; classical slow-rolling was sufficient. See the next section for the overview of this.

\section*{Option (i)}

Having discussed the basic idea, we now revisit the steps taken in~\cite{Berezhiani:2015ola} to estimate the prior probability for a given region to be smooth enough to initiate inflation. In order for a region of size $\ell$ to be well described by a homogeneous configuration of the inflaton field, it must be dominated by $\varphi$-quanta with wavelength longer than $\ell$. Assuming $\la k \ra_{k\ell<1}\simeq \ell^{-1}$ and
$\la k \ra_{k\ell>1}\simeq E$, for the situation of physical interest $\ell^{-1}\ll E$, one arrives at the following requirement
\beq
N_{k\ell<1}\ell^{-1}\gg N_{k\ell>1}E\,, \quad \Rightarrow \quad \frac{N_{k\ell<1}}{N_{k\ell>1}}\gg \sqrt{\ell \mpl}\,.
\label{fraction}
\eeq
Here, in the last step, we have used $E=\rho^{1/4}$ and have required the curvature radius to be shorter than the size of the patch. In other words, most of the $\varphi$-quanta must have wavelengths longer than the size of the patch in order to be able to treat their configuration as a homogeneous classical background\footnote{Note that $\ell\mpl\gg 1$.}. Furthermore, the energy density should be high-enough that the curvature radius is shorter than the size of the region. Considering \eqref{fraction}, the probability of a given patch being dominated by soft quanta is well-approximated by the probability for all quanta within the patch to be soft. Therefore, the probability for a given patch being capable of seeding inflation with curvature scale $H$ can be estimated as\footnote{Strictly speaking, this is the probability for having a homogeneous inflaton configuration. In order for most of its energy to be in the form of slowly-rolling potential energy, the constituent quanta also need to be non-relativistic. We will come to this point later on.}
\beq
\mathcal{P}_{inf}(H)=\mathcal{P}_N\cdot p\left(k<\ell^{-1}\right)^N\, ,
\label{probinf}
\eeq
where $N$ stands for the number of quanta required to drive inflation with curvature scale $H$, within a patch of size $\ell$;
$\mathcal{P}_N$ is the probability of having $N$ quanta within the region in question; and $p(k<\ell)$ is the probability that a given quantum has wavelength longer than the size of this region (given by \eqref{prob1}). It is straightforward to derive a lower bound on the particle number, defined as that value $N_c$ for which inflation takes place with curvature radius precisely equal to $\ell$. Using the Friedmann equation, this value is
\beq
N\geq N_c\equiv 3\ell^2 \mpl^2\,.
\eeq
In other words, $N_c$ is the number of quanta with average energy $\ell^{-1}$ per particle, such that the cumulative energy density sources inflation with curvature radius $\ell$.

Taking into account that $\mathcal{P}_N$ may not exceed unity for obvious reasons, and using \eqref{prob1}, we obtain
\beq
\mathcal{P}_{inf}(N\geq N_c)<\left( \ell E\right)^{-\alpha N}\,.
\eeq

Since $E$ represents the expected energy of a given quantum, and is set by the energy density, we have $N E\ell^{-3}=E^4$, which allows us to recast the above expression into the simple form
\beq
\mathcal{P}_{inf}(N\geq N_c)<N^{-\alpha N/3}\,.
\eeq
Considering that $N\gg 1$, we can conclude that in the initial state the regions capable of initiating inflation are exponentially rare. However, as we have already commented, the exponential volume factor they are rewarded with may be sufficiently large that the late time volume is  dominated by inflated patches. The necessary condition for this to happen is given by
\beq
\mathcal{P}_{inf}e^{3\mathcal{N}}>1\,,
\eeq
with $\mathcal{N}$ denoting the number of e-folds inflation lasted for. Assuming $\frac{\alpha}{3}{\rm ln}N\sim {\rm few}$, we arrive at
\beq
\mathcal{N}>N\geq N_c\,.
\eeq
It is straightforward to show that $\mathcal{N}> N_c$ translates into the following bound on the Hubble scale for the efficient beginning of inflation
\beq
H>H_*\equiv\sqrt{m\mpl}\,.
\eeq
Interestingly, $H_*$ represents the curvature scale for the onset of the self-reproduction regime (i.e., slow-roll eternal inflation, in which the amplitude of quantum fluctuations on horizon scales dominates over the classical variation of the field-value over the Hubble time.) It is worth mentioning that, although this regime is reasonable if it is well-described by the semi-classical picture, there exist arguments suggesting that regions that begin with such a small slow-roll parameter may cease to be described by semi-classical physics before reheating~\cite{Dvali:2013eja,Berezhiani:2016grw}. The essence of the argument is the following. Semi-classically the inflationary background pays no price for producing particles such as, for instance, gravitational waves. However, quantum mechanically it must lose at least one out of the $N$ quantum constituents of the classical background during particle production. Therefore, a configuration with $N$ quantum constituents is expected to expire within $N$ e-folds\footnote{In \cite{Berezhiani:2015ola}, it was quoted from \cite{Dvali:2013eja} that the maximal number of e-folds for which one could trust the semi-classical picture was $N^{2/3}$. Later, in \cite{Berezhiani:2016grw}, the estimate was improved to establish that for inflation $\mathcal{N}_{\rm max}=N$, matching the same result for pure de Sitter space obtained in \cite{Dvali:2014gua,Dvali:2017eba}.}. These argument suggest that option (i) may be an inefficient way of beginning inflation, or that at best it may be sensitive to our understanding of the quantum aspects of eternal inflation.

\section*{Option (ii)}

Now let us turn to the second option for the beginning of inflation. The idea we would like to reconsider is whether it is plausible for an inhomogeneous universe to become homogeneous as a result of a decelerating expansion. One often-discussed possibility, which is supported by numerical simulations for broad but limited sets of initial conditions, is that any inhomogeneities will be eventually washed away, leaving behind the potential energy of the inflaton, which is essentially frozen at curvature scales greater than the inflaton mass. We would like to revisit some aspects of this scenario within the corpuscular description of the initial state, in which different forms of energy correspond to the distribution of energy among quanta of different wavelengths. 

In order to start gaining some intuition, let us consider a Hubble patch filled with inflaton quanta with wavelengths significantly shorter than the curvature radius. (Say, if the average energy density were around Grand Unification scale, $\rho\sim M_{\rm GUT}^4$, then imagine that all $\varphi$ quanta have momenta around $M_{\rm GUT}\gg H_{\rm GUT}$). Let us further assume that on sufficiently large scales, averaging over all components, we have homogeneous decelerated expansion. We may then ask how much energy is in the form of potential energy, which remains essentially unredshifted. Naively, one might be inclined to estimate this as the initial rest energy, in the hope that any kinetic energy would be washed away by the expansion. However, the number density of particles having momenta shorter than the horizon scale is certainly redshifted, so that in this case the potential energy will never come to dominate. Thus, sub-horizon quanta do not contribute to the potential energy budget. Thus, the only quanta that can contribute to the potential energy are the super-horizon ones.

Now, among super-horizon modes, keeping in mind that $m\ll H$, we need not say much about quanta with $k\ll m$, since these definitely contribute to the potential energy because they are practically frozen. The energy density of relativistic super-horizon modes, on the other hand, redshifts until they become non-relativistic\footnote{This is an oversimplification in favour of inflation, as the presence of certain super-horizon perturbations could in fact hinder its onset \cite{Clough:2016ymm}.}.

We begin by estimating the potential energy density budget of a region, within our corpuscular parametrization of the initial state. For this, let us consider a region with relatively homogeneous coarse grained energy density (over scales longer than the average inter-particle separation) $\rho$ and with the curvature scale $H\gg m$. As we mentioned earlier, the modes that are guaranteed to contribute to the potential energy are the non-relativistic modes. To estimate their number, first note (see also the previous section), that since the energy density is given by $\rho\equiv E^4$, and the typical energy of a quantum is $E$, then the expected number of quanta within the Hubble patch is given by
\beq
N=\frac{E^3}{H^3}\ \,.
\eeq
From this, and the crucial property of the distribution function~(\ref{prob1}), we can then estimate the number of non-relativistic quanta as
\beq
N_{\rm soft}=N p(k<m)=\left( \frac{E}{H}\right)^{3} \left(\frac{m}{E}\right)^{\alpha}\,.
\eeq
Using this, the corresponding energy density reduces to
\beq
\rho_{\rm soft}=m H^3 N_{\rm soft} \ .
\label{rhosoft}
\eeq
For $\alpha=1$, the smallest possible value consistent with our assumptions, we finally arrive at
\beq
\rho_{\rm soft}=m^2 E^2\, .
\eeq
Note that higher values of $\alpha$ lead to a stronger suppression of this potential energy density budget.

Assuming that this energy density does not redshift as a result of expansion, it is not possible to show that it is adequate for eventually driving inflation. The reason is simple, the absolute lower bound on the energy density for inflation (within the model at hand) is given by $m^2 \mpl^2$ and consequently requiring that $\rho_{\rm soft }$ exceed this lower bound implies
\beq
E>\mpl\, ,
\eeq
which is outside the validity of the effective field theory.

Having shown that the initial energy stored in non-relativistic quanta is insufficient to provide the required potential energy for inflation, we switch our attention to super-horizon modes that are initially relativistic, but which could eventually contribute to the slowly-rolling potential energy, because of redshift effects. We will refer to these modes as {\it softening} modes. One might imagine that these modes would most likely soon reenter the horizon and then be unable to contribute. However, if the expansion is dominated by negative spatial curvature, then $H\sim a^{-1}$. Therefore, the super-horizon modes will remain super-horizon throughout the expansion, while redshifting. The question we would like to ask is whether these initially relativistic super-horizon modes might contribute an appreciable amount to the potential energy. To examine this, assume an initial Hubble patch with curvature scale $H_0$, filled with $\varphi$ quanta with the properties we have adopted above. As before, their total number is given by
\beq
N=\frac{E^3}{H_0^3}\,.
\label{totnum}
\eeq
Suppose further that we are interested in inflation beginning at the curvature scale $H_{\rm inf}>m$. Then, besides modes that were already non-relativistic at early, pre-inflationary times, modes with
\beq
k\lesssim \frac{H_0}{H_{\rm inf}} m\,
\eeq
will also contribute to the potential energy by the time that the curvature has reduced to $H_{\rm inf}$. We may calculate the total number of such modes using
\beq
dN_k=Nf(k)dk\, ,
\eeq
with $f(k)$ denoting the 1-particle probability distribution that corresponds to \eqref{prob1}.

To be conservative, we assume that the energy density of such modes redshifts as $\rho\propto a^{-2}$ until they become non-relativistic\footnote{If it redshifts more rapidly, our conclusions would be even stronger.}, and that it freezes after that, we can estimate their cumulative energy density by the time the curvature scale reaches $H_{\rm inf}$ as follows
\beq
\rho_{\rm softening}&=&\int_{m}^{\frac{H_0}{H_{\rm inf}}m}dN_k~\frac{kH_0^3}{(k/m)^2}\nonumber \\
&=&\rho_{\rm soft}\cdot \frac{-1+\left(\frac{H_0}{H_{\rm inf}}\right)^{\alpha-1}}{\alpha-1}\, ,
\label{rhosoftening}
\eeq
(for $\alpha \neq 1$) with $\rho_{\rm soft}$ denoting the energy density for the initially non-relativistic quanta given by \eqref{rhosoft}. For the case of least suppressed homogeneous component, $\alpha=1$, the above integral yields
\beq
\rho_{\rm softening}=\rho_{\rm soft}\cdot {\rm ln}\left( \frac{H_0}{H_{\rm inf}} \right)\,.
\eeq
Interestingly, after all the optimistic assumptions, the softening modes in question dominate somewhat over the modes that were non-relativistic from the beginning; namely, the largest possible value of the logarithmic enhancement factor is obtained for $H_0\simeq \mpl$ and $H_{\rm inf}\simeq m=10^{-6}\mpl$, corresponding to ${\rm ln}(H_0/H_{\rm inf})\simeq 6$. Nevertheless, the enhancement is not parametrically significant even for $\alpha=1$. While it is true that for higher values of $\alpha$, the softening modes are increasingly dominant over the initially soft modes, this is of little help because all modes are increasingly suppressed as we increase $\alpha$. For example, even for $\alpha=2$, $\rho_{\rm soft}$ itself is suppressed so much that one requires $E>\mpl$ to ensure enough potential energy for a minimal amount of inflation.

Having demonstrated that our assumptions about the initial state of Universe do not provide sufficient potential energy for the eventual onset of inflation when averaging over large scales, we now ask the question in a way that combines the best of options (i) and (ii). In other words, our analysis of option (ii) has shown the rareness of regions with sufficient potential energy to drive inflation (once other forms are washed away), which begs the question whether this rareness can be compensated for by the exponential volume factor, in the spirit of option (i). For this, we begin with a Hubble patch of curvature $H_0$ and total number of particles $N$ given by \eqref{totnum} and estimate the probability of having $N_{\rm soft}$ modes with $k<m$, keeping in mind that the minimal number of such modes\footnote{We are ignoring the softening modes here for the sake of simplicity.} should be such that their initial energy density is greater than $m^2\mpl^2$. Assuming $N\gg 1$ and $Np$ being finite (with $p$ given by \eqref{prob1}), we can approximate the probability that $N_{\rm soft}$ out of $N$ quanta are nonrelativistic using the Poisson distribution
\beq
P(N_{\rm soft})= \frac{(Np)^{N_{\rm soft}}e^{-{Np}}}{N_{\rm soft}!}\, .
\eeq
This needs to be multiplied by the volume factor ${\rm exp}(3\mathcal{N})$, with number of inflationary e-folds being determined in terms of $N_{\rm soft}$ in a straightforward manner. In particular, having $N_{\rm soft}$ nonrelativistic modes corresponds to their energy density being $N_{\rm soft}mH_0^3$. Assuming that this energy density is constant during the course of the decelerated expansion (an assumption that makes inflation more likely), the number of e-folds can be estimated from $N_{\rm soft}mH_0^3\simeq H_{\rm inf}^2\mpl^2$ and $\mathcal{N}\simeq H_{\rm inf}^2/m^2$ (valid for $\mathcal{N}$ at least a few). This reduces to
\beq
\mathcal{N}=\frac{N_{\rm soft}H_0^3}{m\mpl^2}\,.
\eeq
This makes the minimal required value of $N_{\rm soft}$ to be
\beq
N_{\rm soft}^{min}=\frac{m\mpl^2}{H_0^3}\,.
\eeq
Taking into account that this number must be significantly greater than one and that $Np/N^{min}_{\rm soft}<1$, it is straightforward to show that
$P(N_{\rm soft})\cdot {\rm exp}(3\mathcal{N})$
is exponentially small.

\section*{Conclusions}

We have examined the question of the prevalence of inflation under a concrete set of assumptions about the prior state. Taking into account that the inflaton is an ultralight particle in the early Universe, we have envisioned the pre-inflationary state of these quanta to be in the form of a relativistic gas. This is a particularly fair assumption if they were produced through perturbative decays of high-energy quanta that could have populated the Universe beforehand. Moreover, we have assumed there to be a small probability for a given quantum to be produced soft rather than relativistic. As a result, we expect the existence of rare regions within which inflation may begin, either from the beginning, due to accidentally favorable conditions, or as a result of further evolution and homogenization. Our estimations demonstrate that if the initial state possesses the outlined properties, then the efficient beginning of inflation seems to require access to the part of the potential which is flat enough for hosting slow-roll eternal inflation (a.k.a self-reproduction). 

Our assumptions about the initial state include that the scalar field is initially in the vacuum, corresponding to vanishing energy density, before the perturbative population of the inflaton state began. As such, it is not surprising that we obtained a highly suppressed density in ultra-soft modes (and a correspondingly small contribution to the initial potential energy). 

Alternatively, consider the scenario in which the inflaton potential has degenerate minima. In this case, e.g. for the potential $V(\varphi)=\lambda\left( \varphi^2-v^2 \right)^2$, one usually expects some regions of the universe to be found in one vacuum, and the others in the opposite one, inevitably leaving us with regions in which the scalar field assumes intermediate values. Such intermediate regions, representing topological defects, would not be exponentially rare, and thus our assumptions might require revision in this context\footnote{We would like thank Alex Vilenkin for raising this point.}. Within each region in the same vacuum, at the moment of formation our assumptions regarding the state of $\varphi$ quanta should apply. The only peculiar parts are the domain walls themselves, and their implications are sensitive to their width. If the potential is such that the domain wall is thinner than the curvature radius inside the core, then it will not seed inflation. If its width exceeds the curvature scale, on the other hand, then it would mean that we have sufficiently large regions of homogeneous scalar field inside the topological defects, known as 'topological inflation' \cite{Vilenkin:1994pv,Linde:1994hy,deLaix:1998sf}. Semi-classically these regions then undergo slow-roll eternal inflation. Therefore, we once again  encounter the relevance of this quantum regime.

To summarize, an unequivocally tantalizing idea that makes these issues irrelevant is slow-roll eternal inflation. It is therefore extremely important to understand the true quantum dynamics of that process. Our findings, for the initial conditions at hand, underscore its importance in an interesting way. As we have shown, even if one does not invoke self-reproduction (which yields a practically infinite homogeneous volume) but assumes a simple slow-roll of the inflaton from field values deep inside this quantum regime, most of the late-time volume can be still dominated by inflated patches. However, as we already pointed out, this regime is far from being fully understood. For series of recent papers discussing the issues with the semiclassical picture of this quantum regime see \cite{Dvali:2013eja,Berezhiani:2016grw,Dvali:2017eba}.

\section*{Acknowledgements}

We would like to thank Gia Dvali, Aaron Levy and Eugene Lim for valuable discussions. The work of MT is supported in part by US Department of Energy (HEP) Award DE-SC0013528, by NASA ATP grant 80NSSC18K0694, and by the Simons Foundation Origins of the Universe Initiative.

\end{document}